\documentclass[conference]{IEEEtran}
\usepackage{cite}
\usepackage[english]{babel}
\usepackage[normalem]{ulem}
\usepackage{epstopdf}
\usepackage{times}
\usepackage{mathptm}
\usepackage[pdftex]{graphicx}
\usepackage{subfigure}
\usepackage{algorithmic}
\graphicspath {{./figures}}
\usepackage{balance}
\usepackage{refstyle}
\usepackage[perpage,symbol]{footmisc}
\usepackage{graphicx}
\usepackage{epstopdf}
\usepackage{graphics}
\usepackage{amsmath}
\usepackage{epsfig}
\usepackage{balance}
\usepackage[]{algorithm2e}
\usepackage{algorithmic}
\usepackage{color}
\usepackage{multirow}
\usepackage{cite}

\ifCLASSINFOpdf
\else
\fi

\begin{document}
%
\begin{titlepage}
\title{Deep RNN-Oriented Paradigm Shift through BOCANet: Broken Obfuscated Circuit Attack}
\centering
\end{titlepage}

\author{
	{Fatemeh Tehranipoor$^{\star }$, Nima Karimian$^{\star \star}$, Mehran Mozaffari Kermani$^{\dagger}$, and Hamid Mahmoodi$^{\star }$} \\
            {\normalsize $^{\star }$ School of Engineering, San Francisco State University, USA}\\
            {\normalsize  $^{\star \star}$ Electrical and Computer Engineering, University of Connecticut, USA}\\
            {\normalsize $^{\dagger}$ Computer Science and Engineering Department, University of South Florida}\\
             {\normalsize [tehranipoor@sfsu.edu]}\\
	    \and
  }

\maketitle

\begin{abstract}
This is the first work augmenting hardware attacks mounted on obfuscated circuits by incorporating deep recurrent neural network ($D-RNN$). Logic encryption obfuscation has been used for thwarting counterfeiting, overproduction, and reverse engineering but vulnerable to attacks. There have been efficient schemes, e.g., $satisfiability-checking$ (SAT) based attack, which can potentially compromise hardware obfuscation circuits. Nevertheless, not only there exist countermeasures against such attacks in the state-of-the-art (including the recent delay+logic locking ($DLL$) scheme in $DAC'17$), but the sheer amount of time/resources to mount the attack could hinder its efficacy. In this paper, we propose a deep $RNN$-oriented approach, called $BOCANet$, to (i) compromise the obfuscated hardware at least an order-of magnitude more efficiently ($>20X$ faster with relatively high success rate) compared to existing attacks; (ii) attack such locked hardware even when the resources to the attacker are only limited to insignificant number of I/O pairs ($<$ 0.5\%) to reconstruct the secret key; and (iii) break a number of experimented benchmarks ($ISCAS-85$ $c423$, $c1355$, $c1908$, and $c7552$) successfully.
\end{abstract}

\\

%
\IEEEpeerreviewmaketitle

\section{\textbf{INTRODUCTION}}

Over the past two decades, many of the leading semiconductor companies have become fabless due to the fact that the increasing costs and complexity confine them not to design, test, fabricate, and package ICs. The trend in recent past has been to move towards the globalization of supply chains, and, unfortunately, there exist a number of security threats associated with such exposure. It is well-known that various vulnerabilities introduce numerous opportunities for malicious parties to engage in IP piracy, counterfeiting, and reverse engineering. Hardware obfuscation is a state-of-the-art technique that can be utilized to protect semiconductor IPs at various levels of abstraction \cite{11-Book-obfuscation}.

\begin{figure}
	\centering\includegraphics[width=\linewidth]{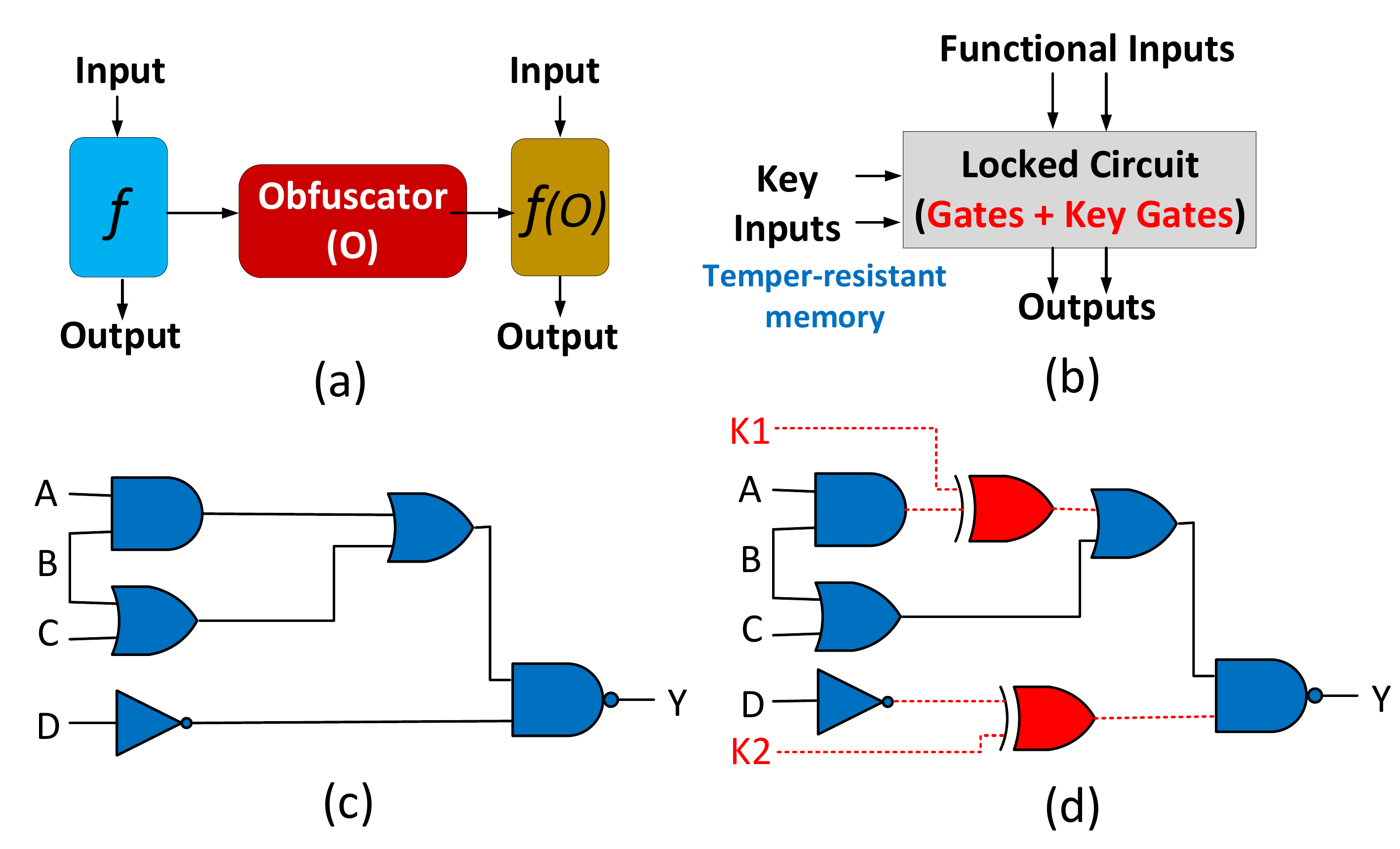}
	\caption{Schematic of (a) a general obfuscation scheme, (b) a general logic locking obfuscation, (c) an original circuit without key gates and key inputs, and (d) an encrypted circuit with key gates and key inputs.}
	\label{logiclocking}
	\vspace{-1.8em}
\end{figure}

Hardware obfuscation technique is the process of transforming a function \textit{F} to another function \textit{F(O)}, which has the same functionality as \textit{F}, e.g., input/output nature, and conceals (locks) the functionality of \textit{F} (as shown in Fig.~\ref{logiclocking}(a)) and/or the structure of an IC from attackers. Through logic encryption, a variant of hardware obfuscation techniques, a given combinational circuit is modified with newly-added inputs, denoted as key inputs, to ensure that the encrypted/locked circuit operates correctly, i.e., produces correct outputs, under a specific key value assertion (valid key). In logic encryption obfuscation, the correct key must not be accessible to the untrusted foundry. Otherwise, the foundry can use it to activate overproduction of ICs and sell them improperly and illegally~\cite{5-Ankur-DAC}. Unfortunately, logic encryption has been recently proven to be compromised by a number of research groups, leaving it potentially vulnerable to various attacks in order to learn the correct key (secret key). Examples of such attacks include satisfiability-checking (SAT) based attack in \cite{4-Malik-HOST} proposed recently by Subramayan \textit{et al.}, a fault analysis attack that directly propagates the key bits to the circuit outputs proposed in \cite{10-Rajjendran-DAC}, and a randomized, local key-searching algorithm to search the key that can satisfy a subset of correct input/output patterns proposed in \cite{12-Plaza-TCAD}. Recent research has also focused on attacks/countermeasures for logic locking \cite{last1, last2, last3}. Nevertheless, we note that although SAT attack can successfully break several existing logic locking techniques, not only there exist countermeasures against such attacks in the state-of-the-art (including the recent delay+logic locking (DLL) scheme in \cite{5-Ankur-DAC}), but the sheer amount of time/resources to mount the attack could hinder its efficacy. Note that SAT attack takes few hours to effectively predict the secret key.

In this paper, we propose a new attack model against logic encryption obfuscation using deep recurrent neural network (D-RNN) denoted as BOCANet: Broken obfuscated circuit attack (see~\cite{Goodfellow} for background). Given a very small number of input/output patterns from the logic encryption obfuscated circuit, this attack enables not only to reconstruct the secret key but also predict the unknown outputs by giving the input patterns of the circuit (without the need of secret key) and vice versa. Based on our results, BOCANet can effectively break logic encryption circuits in order to reconstruct the secret key, achieving relatively high success rate. To the best of our knowledge, this is the first work augmenting hardware attacks mounted on hardware obfuscated circuits using deep learning technique.

Specifically, the major contributions of the proposed work are as follows:
\begin{itemize}
	\item \textit{Broken obfuscated circuit attack, BOCANet}: We introduce BOCANet, a novel attack framework scheme. BOCANet incorporates D-RNN to compromise the secret key from obfuscated hardware at least an order-of-magnitude more efficiently ($>20X$ faster with relatively high success rate) compared to SAT attacks.
	\item \textit{BOCANet strength properties}: In order to objectively measure the success of the proposed attack, we test our BOCANet scheme to break a number of experimented benchmarks (ISCAS-85 c423, c1355, c1908, and c7552) successfully with $32$-bit key size (100\% success rate), 64-bit key size (94\% success rate), 128-bit key size (92\% success rate), and 256-bit key size (89\% success rate), respectively.
     \item \textit{BOCANet obliviousness of secret key}: Through our experiments, we show that BOCANet is oblivious of the secret key such that given a number of trained input/output patterns to D-RNN, it is capable of predicting any unknown I/O pairs without the knowledge of circuit's secret key.
\end{itemize}

We first present the preliminaries in Section 2. In Section 3, we present the proposed scheme, BOCANet. Section 4 presents the results of our proposed attacks assessments on logic locking obfuscated circuits along with evaluating four ISCAS-85 benchmarks. We conclude the paper in Section 5.

\section{\textbf{PRELIMINARIES}}

\subsection{\textbf{Logic-Based Hardware Obfuscation}}
Logic-based hardware obfuscation (or logic encryption, logic locking) is an emerging and promising technique to thwart the threat of counterfeiting, overproduction, and reverse engineering by an untrusted foundry \cite{1-bhunia-DAC, 2-karri-DAC, 3-farinaz-DATE}. Through logic locking, a given combinational circuit is modified with newly-added inputs to ensure that the encryption circuit produces correct outputs under a specific key value assertion (valid key). Upon applying a wrong key, the encrypted design will exhibit a wrong functionality, i.e., produces wrong outputs. Since the design is encrypted by the designer, the foundry cannot use any copies or overproduce ICs without the secret keys. Logic encryption hides the functionality and the implementation of a design by inserting additional gates into the original design. We note that the gates inserted for encryption are key-controlled gates. Fig.~\ref{logiclocking}(b) shows a general schematic of logic locking obfuscation technique which contains functional input (input patterns to the circuit), key input (secret key that is stored in tamper resistant memory), locked circuit (including normal gates and added gates for key gates), and outputs from the locked circuit. Fig.~\ref{logiclocking}(c) and Fig.~\ref{logiclocking}(d) illustrate a circuit without/with key gates, respectively (modulo-2 addition through XORing with K1 and K2).

\subsection{\textbf{SAT Attacks}}
Satisfiability-checking (SAT) based attack is a newly-proposed attack \cite{4-Malik-HOST}, threatening the security of logic encryption circuits by deciphering a functionality-correct key (w.r.t. combinational logic) of most locking techniques within a few hours \cite{5-Ankur-DAC}.  The main step in the SAT-based attack is to use two copies of the encryption logic circuit with the same input but different keys (lock-keys), under a given constraint, to check whether it is still possible to generate different outputs from the locked circuit. Such small number of input/output pairs are denoted as differentiating input patterns (DIPs) that can be used to eliminate incorrectly-guessed keys. Each DIP is then used to query the original circuit blackbox to get the correct output. The DIP with output is then used to further constrain the keys under consideration. In fact, the idea of using DIP is to exclude at least one wrong key from consideration. Finding DIPs requires a sequence of SAT formula that can be solved by SAT solvers. In order to defeat SAT attacks, a number of research works have proposed inserting additional SAT attack resistant logic blocks, e.g., the Anti-SAT block by Xie \textit{et al.} \cite{6-Ankur-CHES} and SARLock by Yasin \textit{et al.} \cite{7-Yasin-HOST}.

AppSAT is another attack to logic locking which has been presented in HOST 2017. AppSAT~\cite{shamsi2017appsat} is a technique for approximate deobfuscation based on the active learning (semi-supervised machine learning) with random querying and intermediate error estimation. The main goal in active-learning is to find a querying strategy that minimizing the number of queries required to learn the target function. While active learning is sample-efficient, it can be computationally expensive since it requires iterative retraining. To speed this up, we propose a lightweight architecture based on LSTM. Our model is not only computationally much more efficient than AppSAT, but also its accuracy is higher than that of AppSAT mechanism can provide. Another Advantage of our work compare with other attacks is that when a different key is applied to the same benchmark, deep learning does not need to be trained from scratch one more time. Consequently, we can apply transfer learning by taking a fully-trained model for the existing BOCANet and retrain from the existing weights for new benchmark to deobfuscate less than a minutes. The memory usage is remained below 1GB with a quarter of our CPU usage.

\section{\textbf{OVERVIEW OF THE PROPOSED BOCANET SCHEME}}

\subsection{\textbf{Data Acquisition}}
We consider open source obfuscation benchmarks \cite{8-benchmark} for our deep learning-based reverse engineering experiments. These benchmarks are combinational logic circuits with complexities ranging from less than a thousand to more than five thousands gates. These benchmarks utilize the logic-based obfuscation technique using random key gate insertion with key sizes ranging from $32$ bits to $256$ bits. We note that these keys are typically programmed in an on-chip non-volatile memory in a trusted environment after manufacturing.  The attack model assumes that there is access to a fabricated chip with programmed keys. The size of the key is known information as it is visible form the circuit interface, but the key value is unknown. With access to a chip with programmed key, an electrical test can be performed to extract an input/output stimulus/response table. The unknown part is the secret key, and the goal of the deep learning algorithm is to find the key from a given input/output stimulus/response table.

To generate input/output stimulus/response table, we setup a SystemVerilog testbench \cite{9-systemverilog} for each benchmark to simulate its gate level netlist with randomly-generated input stimuli and obtain corresponding output responses. For large number of inputs, it would be prohibitive to simulate all possible input combinations as we would run out of either time and/or computer memory space to generate and store all combinations. Hence, we generate as many random stimuli as sufficient for the deep learning algorithm to come to a reasonable estimate of the key. The accuracy of the key estimate is improved by providing more input/output stimulus/response data.  For the purpose of this experiment, we manually set a key for a design under test, and check to see if the deep learning algorithm can determine the key. We would like to emphasize that this does not confine the proposed scheme, and the proposed approach in this paper is oblivious of the chosen key.

\begin{figure}
	\centering
\includegraphics[width=1\linewidth]{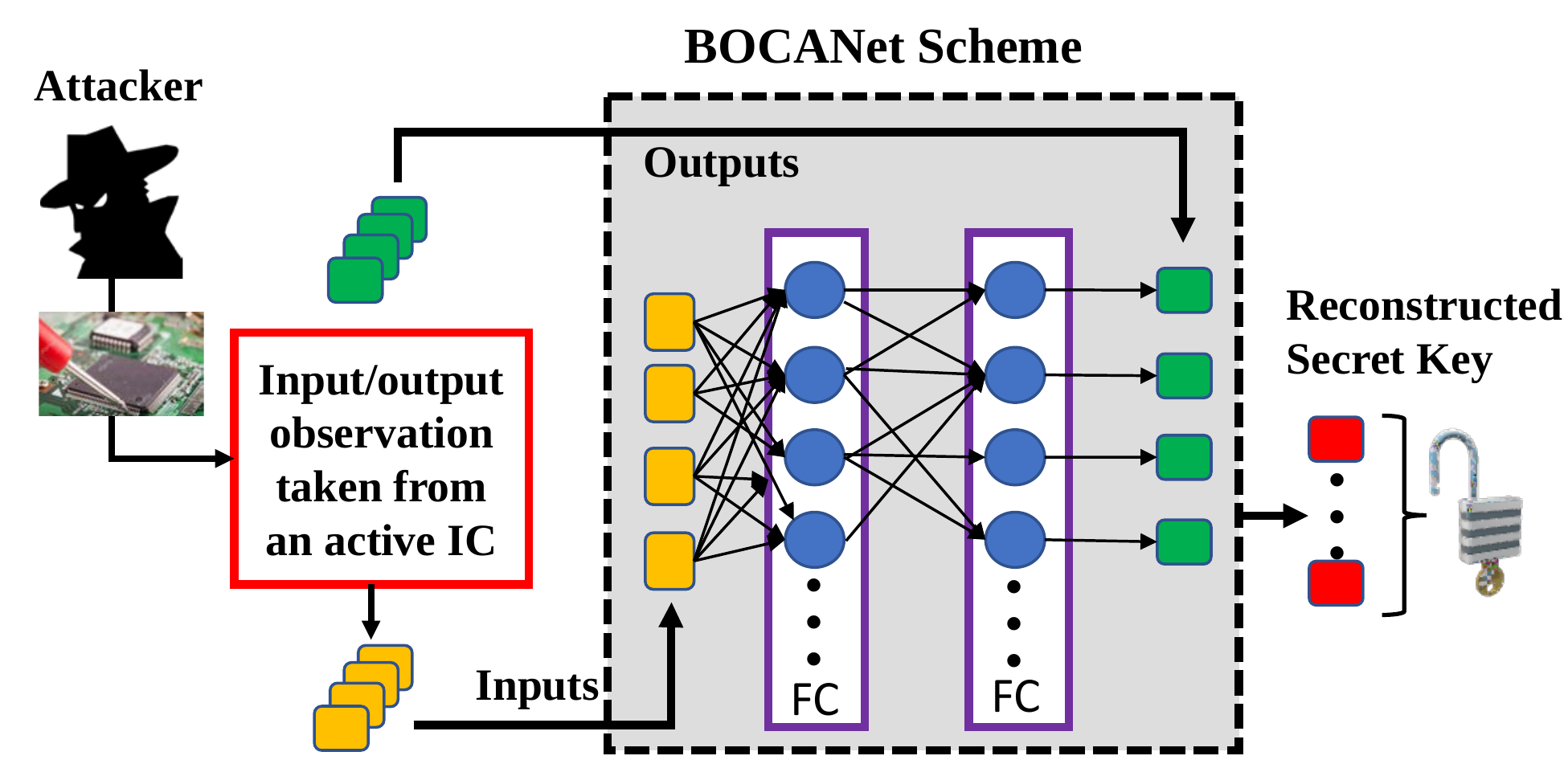}
	\caption{Block diagram of our proposed BOCANet scheme.}
	\label{scheme}
	\vspace{-1.8em}
\end{figure}

\subsection{Attack Scheme}
 In this work, we introduce a new attack called BOCANet attack in which we use deep learning through recurrent neural network, i.e., D-RNN, to predict the secret key. The D-RNN is a deep learning-based algorithm that can perform the same task for every element of a sequence (data), with the output being dependent on the previous computations. In the BOCANet attack, our assumption is that there exist some input/output patterns for an attacker (from an obfuscated circuit) in order to predict the secret key. For these experiments, we have considered four benchmarks \cite{8-benchmark} with different key sizes ($32$-bit, $64$-bit, $128$-bit, and $256$-bit).  Here, we also assume that an attacker has access to only less than 0.5\% of the total  I/O pairs available from the obfuscated circuit.  Fig.~\ref{scheme} illustrates our proposed BOCANet attack model to obfuscated circuits using deep learning techniques. As shown in this figure, an attacker may have access to a set of I/O pairs of an obfuscated circuit by probing (or somehow accessing) them. By knowing the correct I/O pairs, deep learning is able to train itself in order to predict a secret key that has been implemented in the obfuscated hardware. In order to train our BOCANet scheme, we have implemented deep RNN~\cite{Alex} with long short-term memory (LSTM)~\cite{Sutskever}. More details on our BOCANet approach is discussed in the following.

\begin{figure}
	\centering
\includegraphics[width=0.75\linewidth]{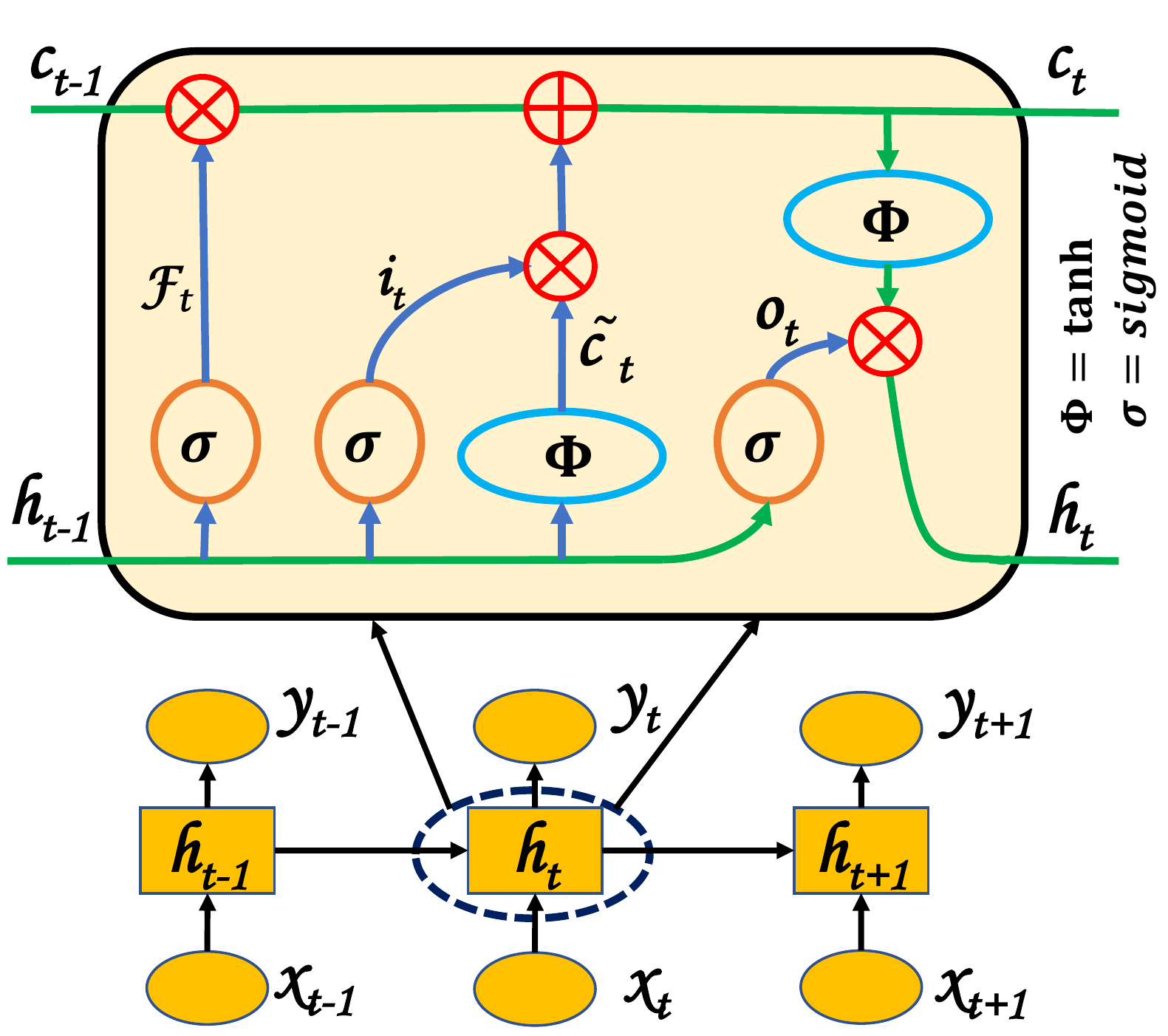}
	\caption{Block diagram of deep RNN with long short-term memory (LSTM).}
	\label{lstm}
	\vspace{-1.8em}
	
\end{figure}

\subsection{\textbf{Deep RNN Implementation}}
Recurrent neural networks have recently attracted prominent attention for modeling variable length sequences~\cite{Hinton}. A D-RNN is a deep structure of a neural network that operates in time. At each time step, it accepts an input vector, updates its hidden layer through non-linear activation functions, and uses it to predict its output. Since D-RNN's hidden layer can store high-dimensional information, and its nonlinear dynamics can implement a powerful computation, it forms D-RNN to form a strong model to perform modeling and prediction functions with a high complex structure. In this paper, we explore deep extension of basic RNN to model our BOCANet for reconstructing secret key from obfuscated circuits. Given a sequence of inputs $x_{1}, x_{2}, x_{3}, ... , x_{p}$ and outputs $y_{1}, y_{2}, y_{3}, ... , y_{q}$, we want to predict secret-key $k_{1}, k_{2}, x_{3}, ... , k_{n}$. In order to do that, we train I/O pairs with LSTM RNN architecture.

LSTM unit refers to a specific architecture of RNNs, introduced by Hochreiter \& Schmidhuber \cite{13-Hochreiter}, targeting to tackle long-term dependencies challenge unsolved in earlier RNN architectures. When learning time-series data, RNNs aim to learn the patterns repeatedly happened in the past by sharing the states that are decomposed into multiple layers in order to gain properties from `deep' architectures~\cite{Sutskever}.  Fig. 3 presents a D-RNN with LSTM cell architecture. LSTM cells have a special sharing parameter vector called memory parameter vector $c_{t}$ deployed to keep the memorized data. In each time stage, the memory parameter has three operations: (i) Discarding useless data from memory vector $c_{t}$, (ii) adding new data $i_{t}$ selected from input $x_{t}$ and previous sharing parameter vector $h_{t-1}$ into memory vector $c_{t}$, and (iii) deciding new sharing parameter vector $h_{t}$ from memory vector $c_{t}$. As shown in the LSTM cell, the sharing memory parameters, $h_{t}$, are passing through various time stages only with two operations to \textit{memorize} new data and \textit{forget} time-out memories. Thus, the sharing memory can conduct useful data for an adequately-long time and results in enhancing the RNN performance:
\begin{align}
\small
\begin{cases}
f_{t}\leftarrow \sigma_{g}(W_{xf}[h_{t-1},x_{t}]+V_{cf}c_{t-1}+b_{f})\\
i_{t}\leftarrow \sigma_{g}(W_{xi}[h_{t-1},x_{t}]+V_{ci}c_{t-1}+b_{i})\\
o_{t}\leftarrow \sigma_{g}(W_{xo}[h_{t-1},x_{t}]+V_{co}c_{t-1}+b_{o})\\
c_{t}\leftarrow f_{t}*c_{t-1}+i_{t}*tanh(W_{xc}[h_{t-1},x_{t}]+b_{c})\\
h_{t}\leftarrow o_{t}*tanh(c_{t})
\end{cases} \label{lstm}
\end{align}  where $f_{t}, i_{t}, o_{t}, c_{t}$, and $h_{t}$ are the forget gate, input gate, output gate, cell, and hidden state output, respectively.

The backpropagation algorithm is used to train the LSTM net, i.e., weights are updated based on the gradient descent error of the output. According to the gradient descent rule, the weights are modified towards minimizing the square error of the output:
\begin{equation}
E\leftarrow\sum_{q=1}^{P}{({D_q}-{Z_q})^2}+ E
\label{equ.}
\end{equation} where $Z_{q}$ is the predicted output of the $q$-th residue in the training
dataset, $D_{q}$ is the target output vector, and $p$ is the number of training patterns. The learning procedure is carried out by updating the weights $W$ proportional to the gradient descent of the error for every training pattern as follows:
\begin{equation}
\delta_{zq}:(d_{q}-z_{q})z_{q}(1-z_{q})  \quad  \quad    q=1,2,...,P
\label{equ.}
\end{equation}

\begin{equation}
\delta_{yj}:y_{j}(1-y_{j})\sum_{q=1}^{P}\delta_{zq}W_{iq} \quad  \quad   j=1,2,...,n
\label{equ.}
\end{equation}

\begin{align}
\small
\begin{cases}
\Delta{W_{jq}}\leftarrow\eta{y_{j}}\delta_{zq}+\alpha\Delta{W_{jq}} \\
W_{jq}\leftarrow{W_{jq}} + \Delta{W_{jq}}
\end{cases} \label{second}
\end{align}
where $\eta$ is called the constant learning rate. To sustain the effect of the past input patterns, the momentum term $\alpha$ is always appended and the weights are optimized through (\ref{second}). The error vector $\delta_{zq}$ is propagated to the back layers and the weights are moderated based on the back propagated error. The constant learning rate and the momentum term are adjusted to $0.01$ and $0.9$, respectively. The network weights are initialized with small random values within $[-0.05, 0.05]$ interval.
\begin{figure}
        \centering
\includegraphics[width=0.8\linewidth]{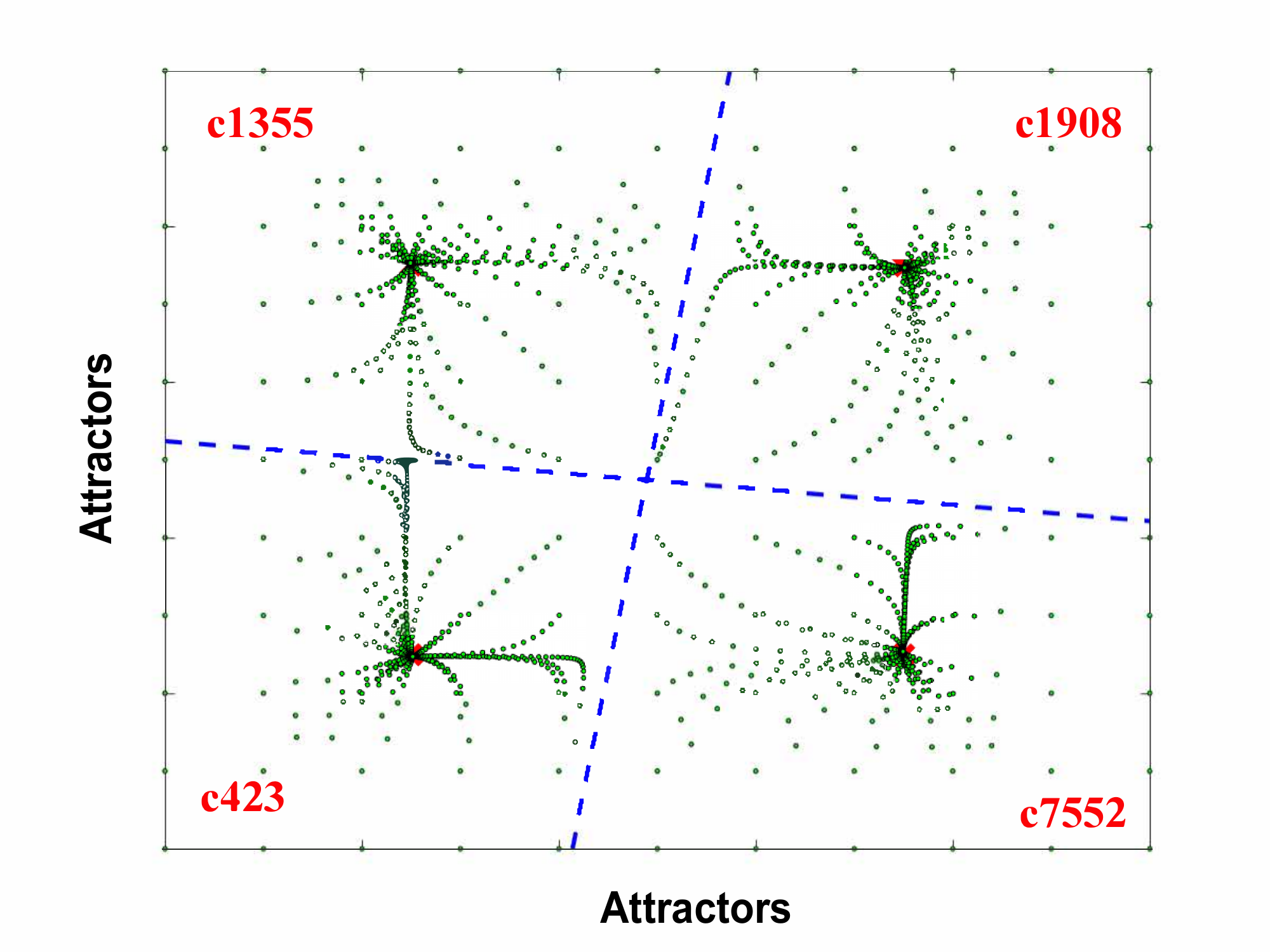}
	\caption{Neuron attractors under 4 different benchmark secret keys.}
	\label{attractor}
	
\end{figure}

\begin{figure}
\begin{tabular}{cc}
\centering
\includegraphics[width=0.45\linewidth]{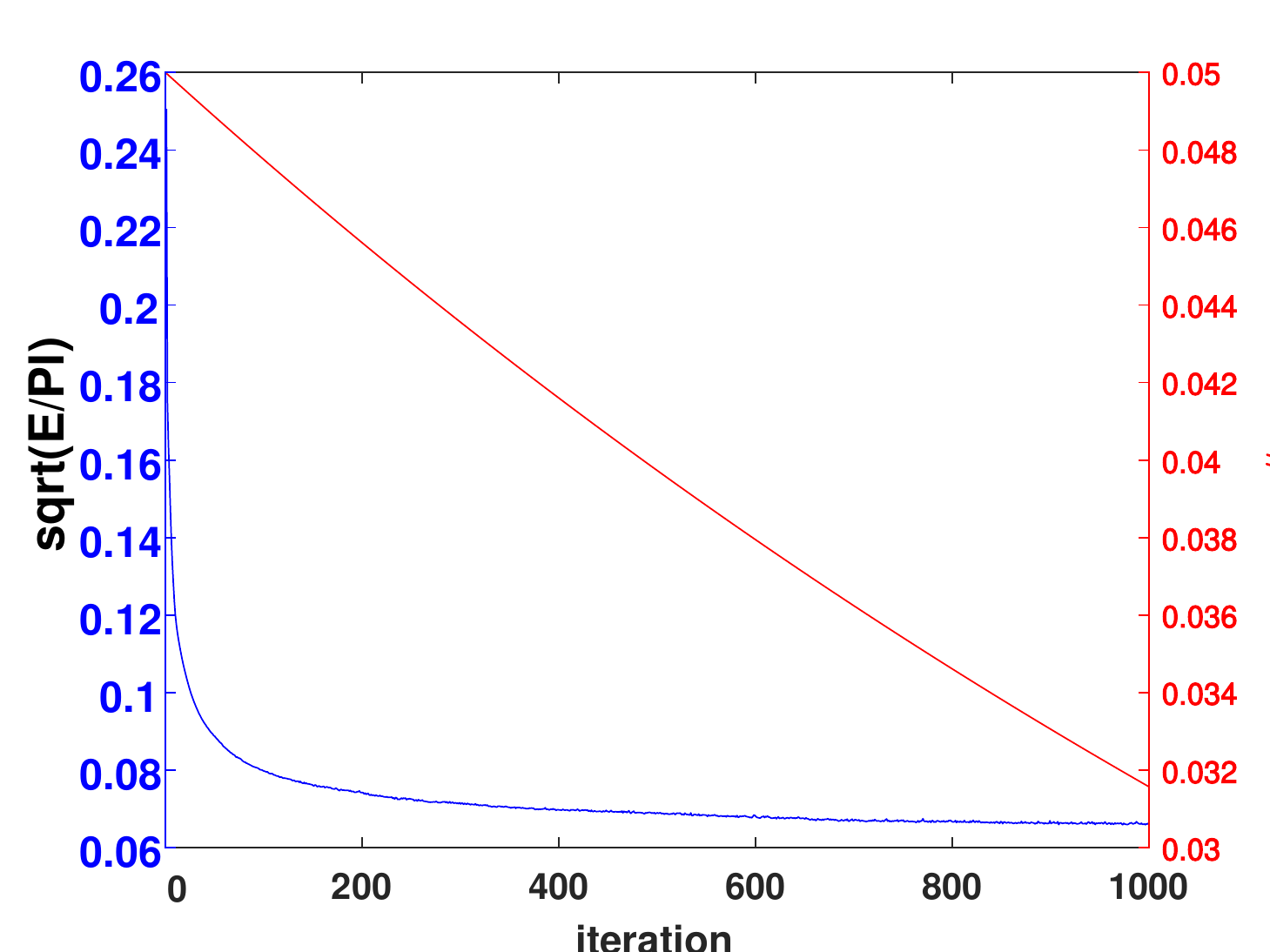} &
\includegraphics[width=0.45\linewidth]{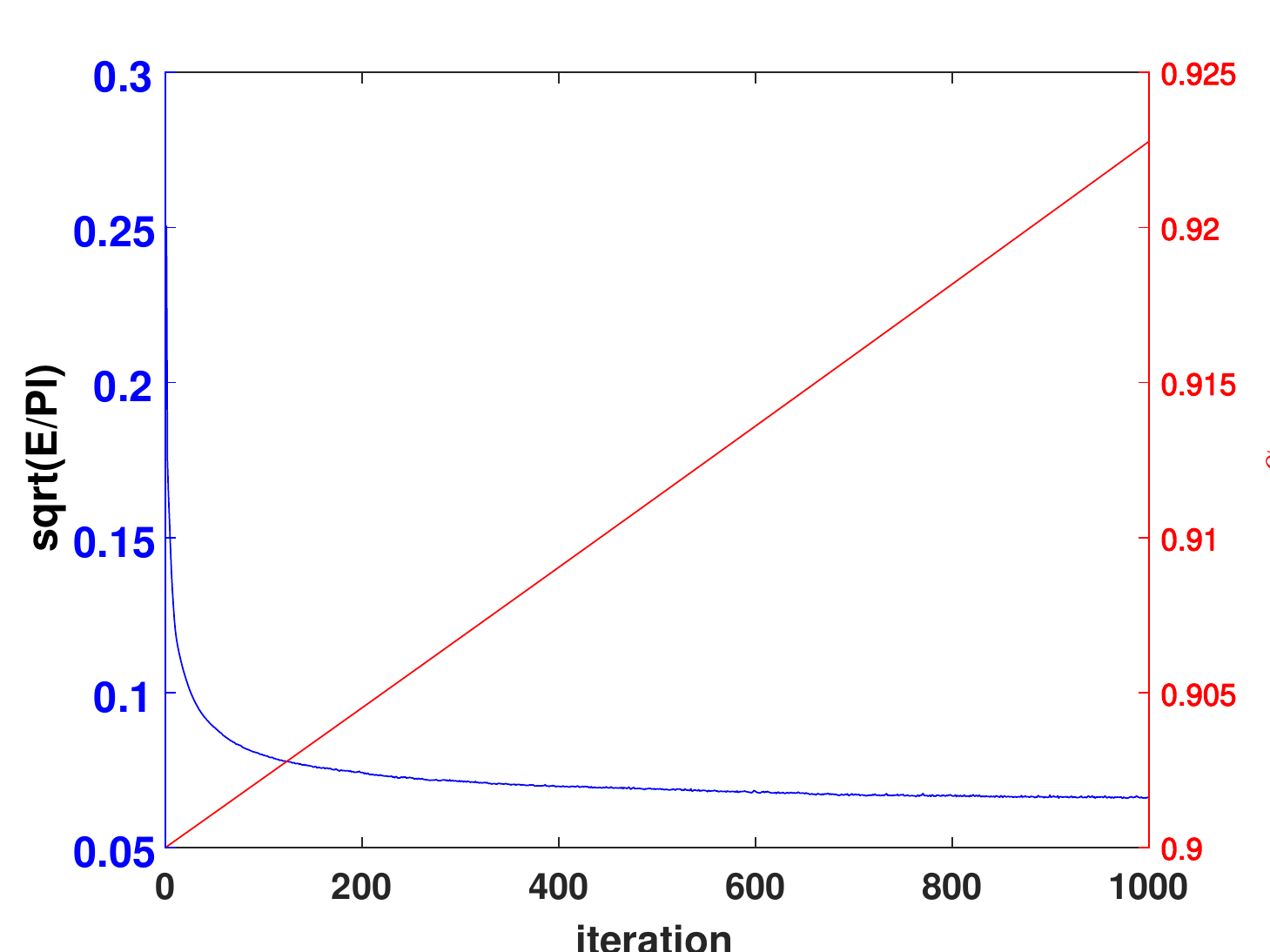} \\
(a) & (b)
\end{tabular}
\caption{The MSE error (blue curve) \textit{vs.} (a) training step (red line) and (b) momentum (red line).}
\label{errors-graph}
\end{figure}

Fig.~\ref{attractor} demonstrates the training process of experimented ISCAS-85 benchmarks in D-RNN in order to reconstruct the secret key. As shown in the figure, there are numerous number of green circle points (nodes) that indicate neuron attractors. These nodes try to settle to a stable pattern (here referred to as secret key) which is shown in the figure by 4 red cross points ($\times$). More precisely, these attractors network  are a set of $N$ network nodes from each benchmark connected in such a way that their global dynamics become stable to predict the secret key from the obfuscated circuits. Furthermore, in order to evaluate the performance of our attack model (BOCANet), one potential metric is mean-square-error (MSE). MSE is a cost function that measures the average of the squares of the errors; the difference between the estimator, i.e., obfuscated circuit output, and what is estimated, i.e., predicted output. Fig.~\ref{errors-graph}(a) illustrates the impact of training step \textit{vs.} momentum coefficient on MSE. As can be seen in Fig.~\ref{errors-graph}(a), by decreasing the training step in each epoch, the D-RNN weight can be adjusted more efficiently than constant training. Fig.~\ref{errors-graph}(b) represents how an increase in momentum has impact on the error rate of trained weights.

\section{\textbf{EXPERIMENTAL RESULTS}}
To break logic encryption obfuscated implementation, we have conducted 3 different DL-based attacks: i) Attack1-Key Reconstruction: In this scenario, the input/output pairs collected from several ISCAS-85 benchmarks from \cite{8-benchmark} are available and the aim is to reconstruct the secret key; ii) Attack2-Output Guessing: In this attack, the attacker applies a set of inputs to deep learning in order to figure out the corresponding outputs (unknown outputs); and iii) Attack3-Input Guessing: In this attack model, the attacker tries to guess the input patterns (unknown inputs) while the outputs are given to the deep learning infrastructure. As indicated earlier, Attack2 and Attack3 are independent from secret key. In other words, they do not require key for prediction of I/Os. We assume that the attacker has access to a set of I/O pairs and the key is unknown to him/her.

\subsection{\textbf{Training BOCANet}}
To train our BOCANet scheme, we consider a small number of input/output observations taken from an activated IC from a huge pool of all possible I/O patterns. For instance, if a benchmark has $N$ inputs, $2^N$ different input stimuli are possible. We assume that the attacker has access to a limited and insignificant number of I/Os (less than 0.5\% of the data). Deep recurrent neural network is trained by \textit{load profile batches} randomly fetched from the load profile pool so that D-RNN is not only learning individual load patterns but also the common sharing load features and uncertainties. All layers of the networks, except for the last ones which contain linear neurons, are made of sigmoid type neurons and have a neuron for bias. We adjust the weights of the neural network to minimize the MSE on training set. In order to prevent local minimum and overfitting, recursive weight coefficients within a layer of back-propagation are chosen as appropriate batch size. In fact, in each of D-RNN's training iterations, the training batch is firstly fetched from the data pool, and then fed into the D-RNN network. Each training batch includes two matrices with fixed size, i.e., input matrix with size $B\times I$ and output matrix with size $B\times O$. The time-cost and iteration of training process highly depend on the feed-in data sequence size \textit{I}, the choice of optimizer, the network size (\text{I}, \textit{H}), and the training batch size \textit{B}. We do not employ a pre-training step; deep architectures are trained from scratch with the supervised mean square error of network. Additionally, we operate early stopping: Out of all iterations, the model with the best development set performance is picked as the final model to be evaluated.

\begin{table}[]
\centering
\caption{Experimented obfuscated ISCAS-85 benchmarks, their characteristics, and the required compromise time}
\label{tablebenchmark}
\begin{tabular}{l|c|c|c|c|}
\cline{2-5}
 & \multicolumn{1}{l|}{\# Inputs} & \multicolumn{1}{l|}{ \# Outputs} & \multicolumn{1}{l|}{Key-size} & \multicolumn{1}{l|}{Attack Time} \\ \hline
\multicolumn{1}{|l|}{\textbf{c423}} & 32 & 7 & 32 Bits & \textbf{6 mins} \\ \hline
\multicolumn{1}{|l|}{\textbf{c1355}} & 41 & 32 & 64 Bits & \textbf{11 mins} \\ \hline
\multicolumn{1}{|l|}{\textbf{c1908}} & 33 & 25 & 128 Bits & \textbf{19 mins} \\ \hline
\multicolumn{1}{|l|}{\textbf{c7552}} & 207 & 108 & 256 Bits &  \textbf{35 mins} \\ \hline
\end{tabular}
\end{table}

\subsubsection{\textbf{Attack1-Key Reconstruction}}
\begin{figure}
	\centering
	\includegraphics[width=0.8\linewidth]{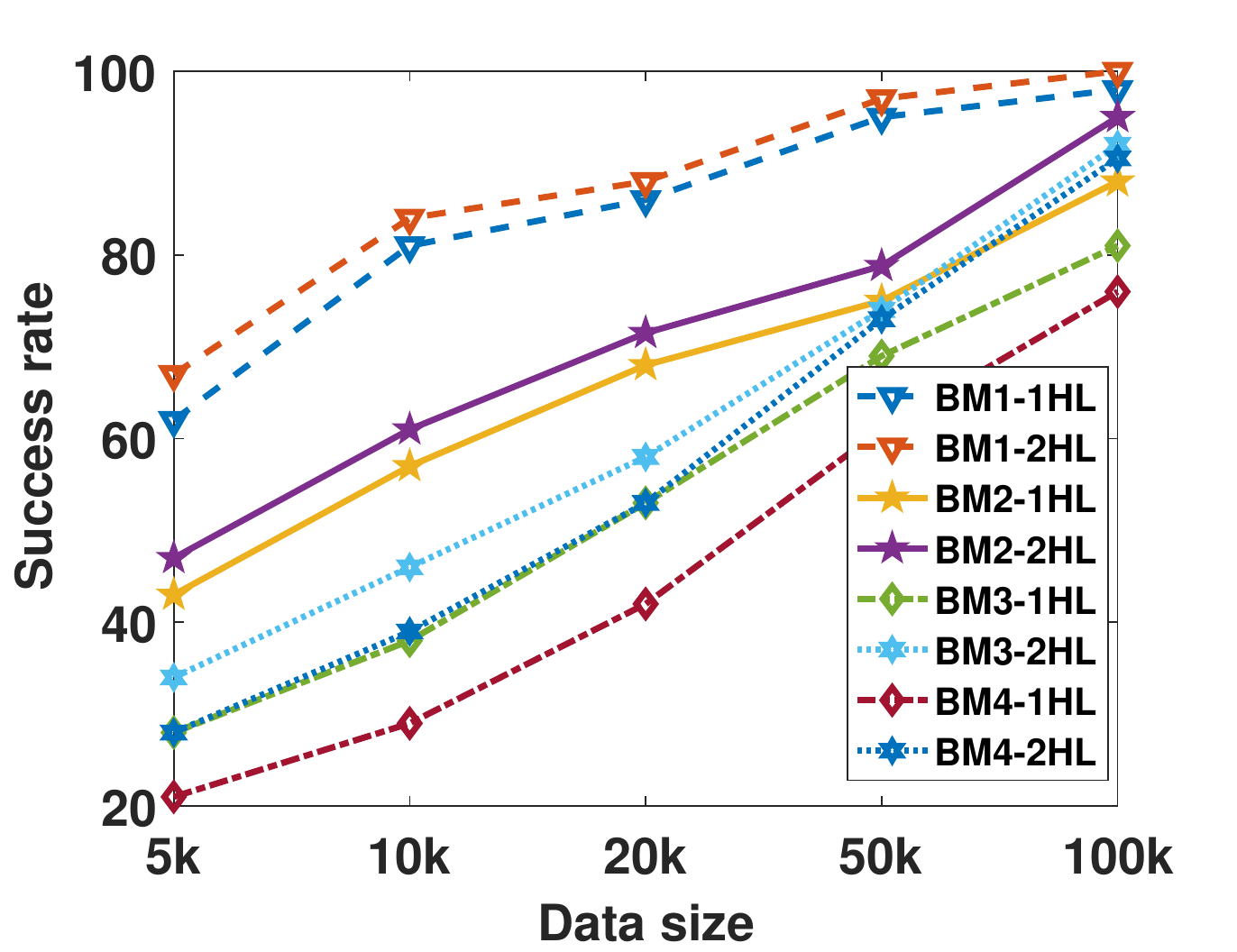}
	\caption{Success rate of BOCANet attack for 4 benchmarks based on one and two hidden layers.}
	\label{accuracyplot}
	\vspace{-1.8em}
\end{figure}

In this attack model, the attacker's objective is to reconstruct the secret key from logic encryption obfuscation circuit. As indicated earlier, using D-RNN, two phases are needed for this attack model. In the training phase, we train input and output pairs without providing any key to neural network. Once D-RNN's parameters and weights are trained, we apply key as an input value. Here, initially, the key value is assigned randomly through BOCANet. Next, during the test phase, the initial key value will be later updated based on the MSE of the trained outputs and the newly-generated outputs due to the presence of secret key in the neural network. We have evaluated this attack on 4 different ISCAS-85 benchmarks with different size of keys (32-bit, 64-bit, 128-bit, and 256-bit). The information regarding the benchmarks are shown in Table~\ref{tablebenchmark}. As seen in this table, we were able to reconstruct the secret key (32 bits) from benchmark ISCAS-85 c423 in 6 minutes from BOKANet \textit{that is at least 20 times faster than SAT attack}. For the second benchmark (ISCAS-85 c1355) which has the key size of 64 bits, prediction duration for the key took only 11 minutes. ISCAS-85 c1908 was the third benchmark that we have tested for this attack. This benchmark has the key size of 128 bits and our BOCANet scheme needed only 19 minutes to break this 128-bit secret key. The last benchmark (ISCAS-85 c7552) has key size of 256 which is very large and, again, our proposed attack using D-RNN could break the key in almost 35 minutes. This benchmark has much higher number of inputs and outputs and key size compared to the other 3 benchmarks. Still, our result is much better than that of SAT attack which takes a few hours to break reasonably-large key sizes. Fig.~\ref{accuracyplot} shows the tradeoff between the training size and successful rate of guessing secret keys for each benchmark. As can be seen, by increasing the training size to 100K, the successful rate goes up. Note that we applied BOCANet with one and two hidden layers of D-RNN which is shown in the figure as "1HL" and "2HL", respectively. As seen in Fig.~\ref{accuracyplot}, the success rate of guessing secret keys (using two hidden layers) for logic encryption obfuscation benchmark1 is 100\%, benchmark2 is 94\%, benchmark3 is 92\%, and for benchmark4 is 89\%. As a result, by having two hidden layers, we achieved higher success rate.

\begin{figure*}
\begin{tabular}{cccc}
\includegraphics[width=0.23\linewidth]{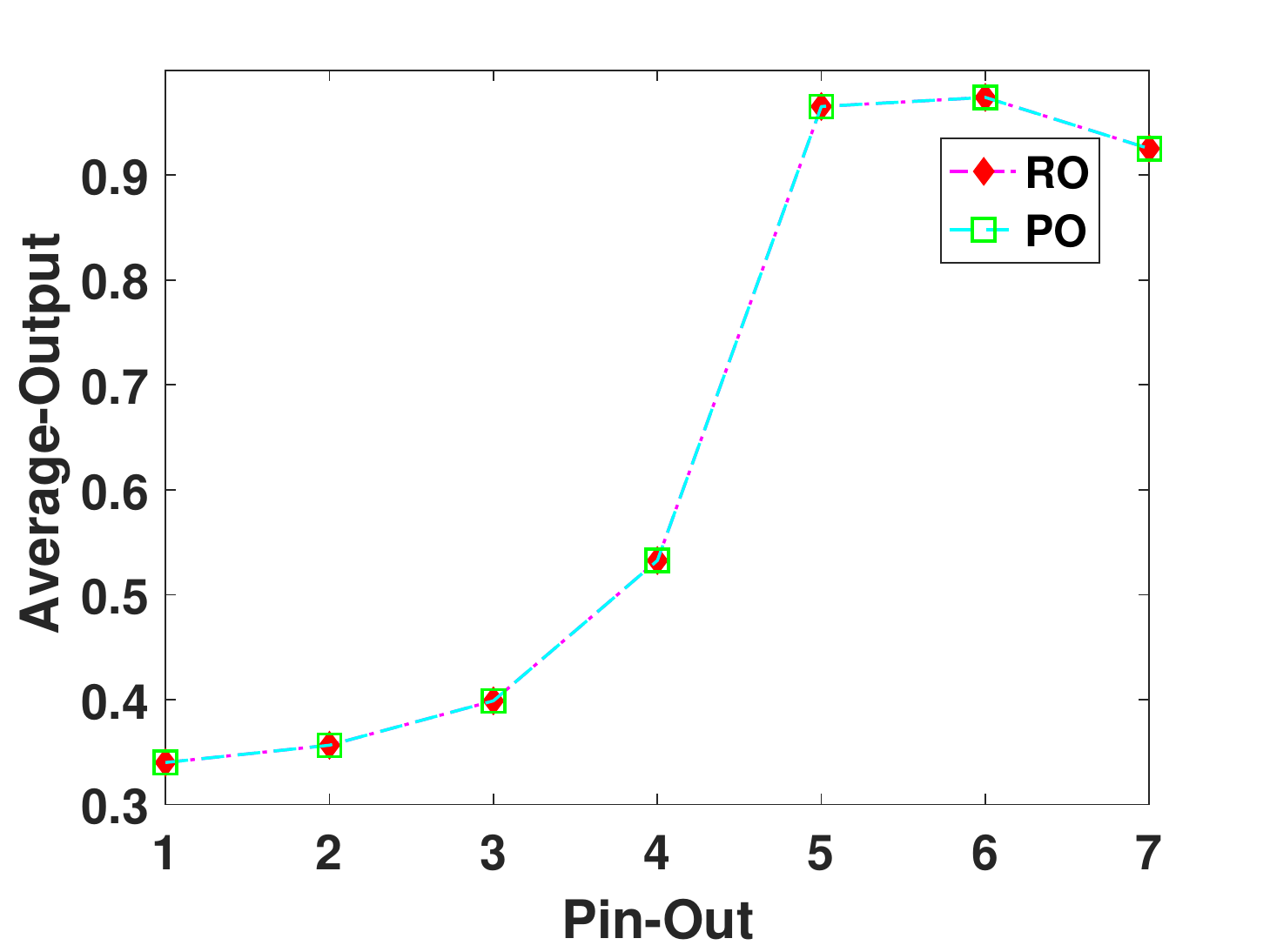} &
\includegraphics[width=0.23\linewidth]{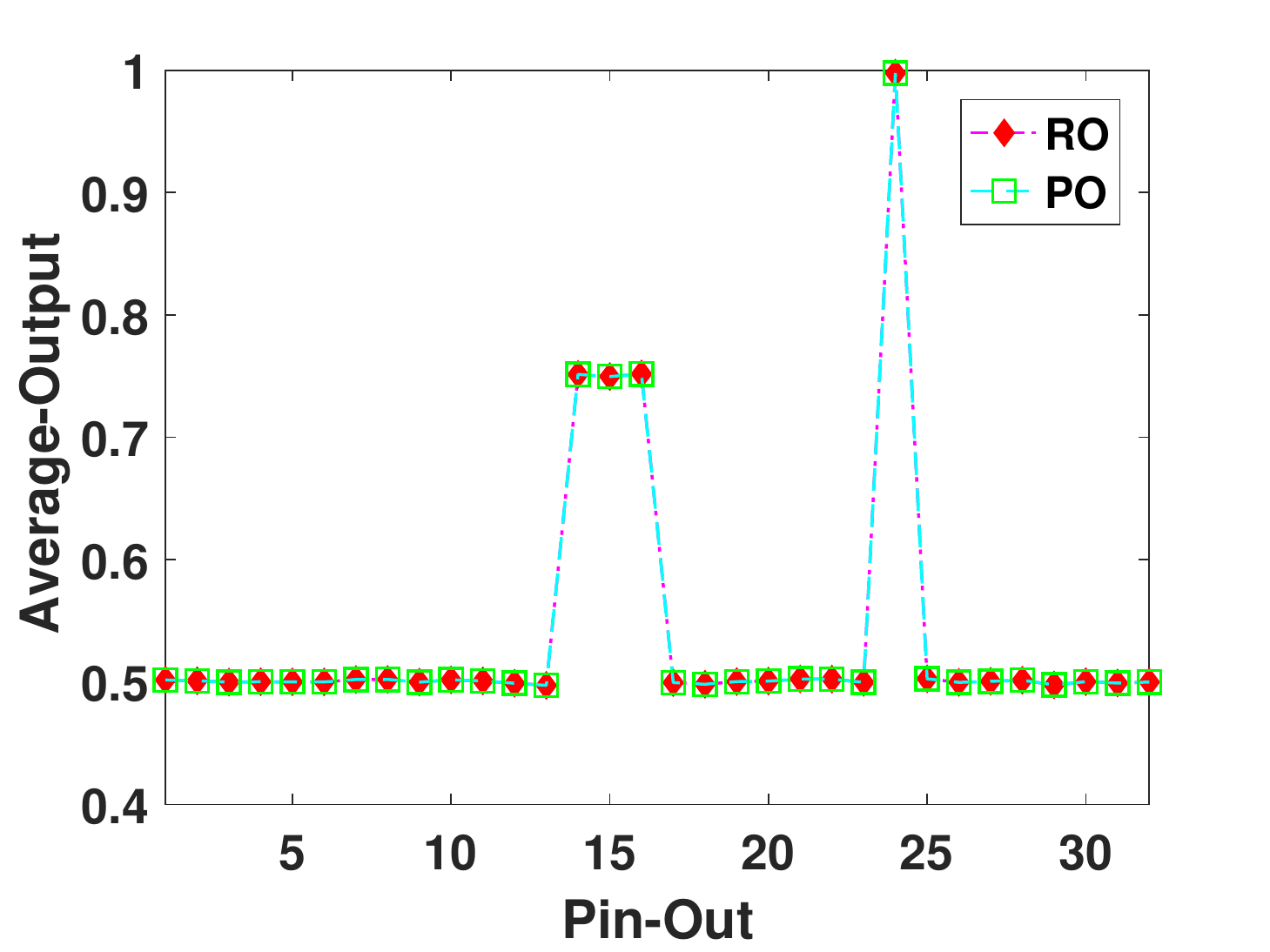} &
\includegraphics[width=0.23\linewidth]{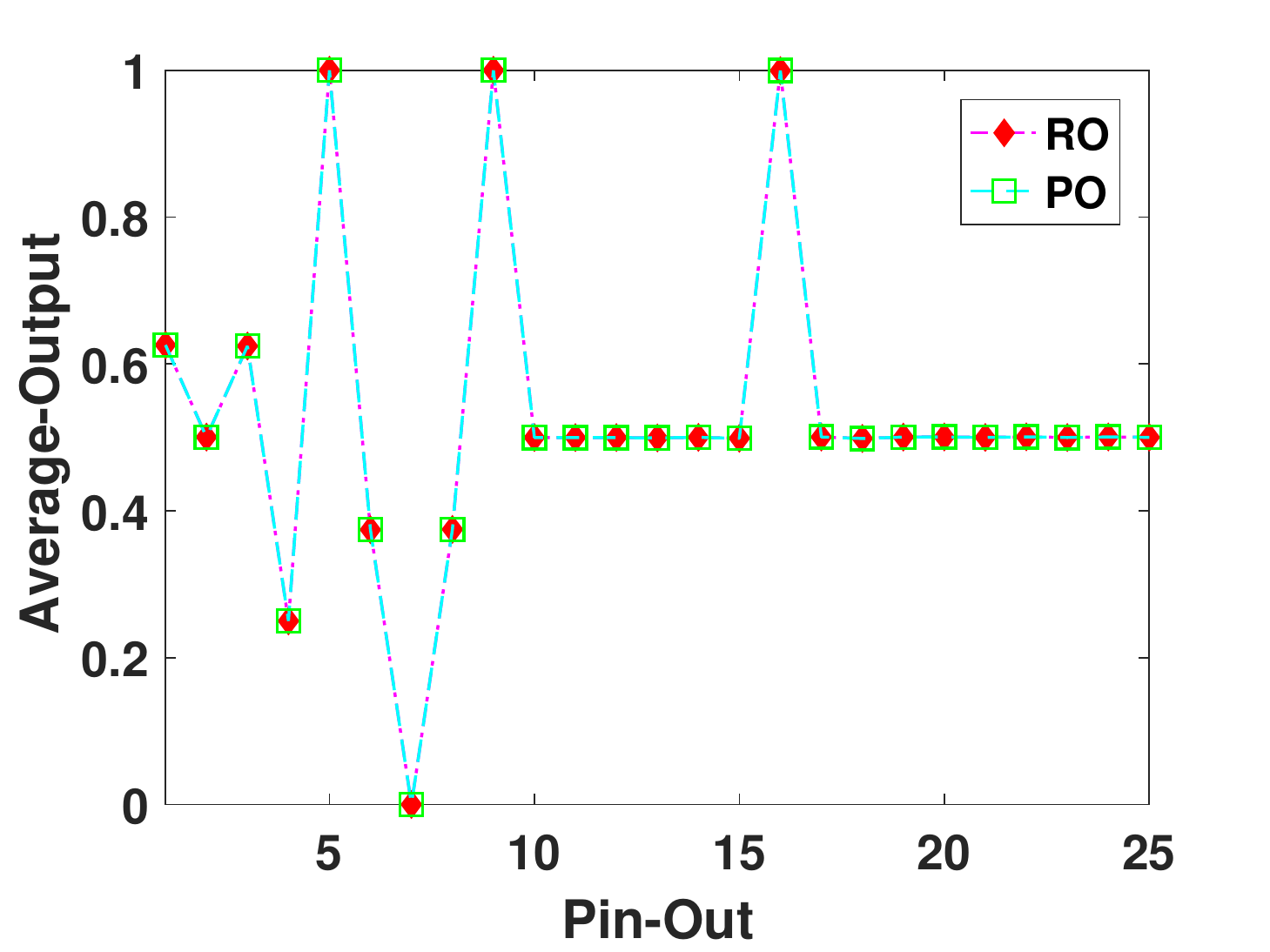} &
\includegraphics[width=0.23\linewidth]{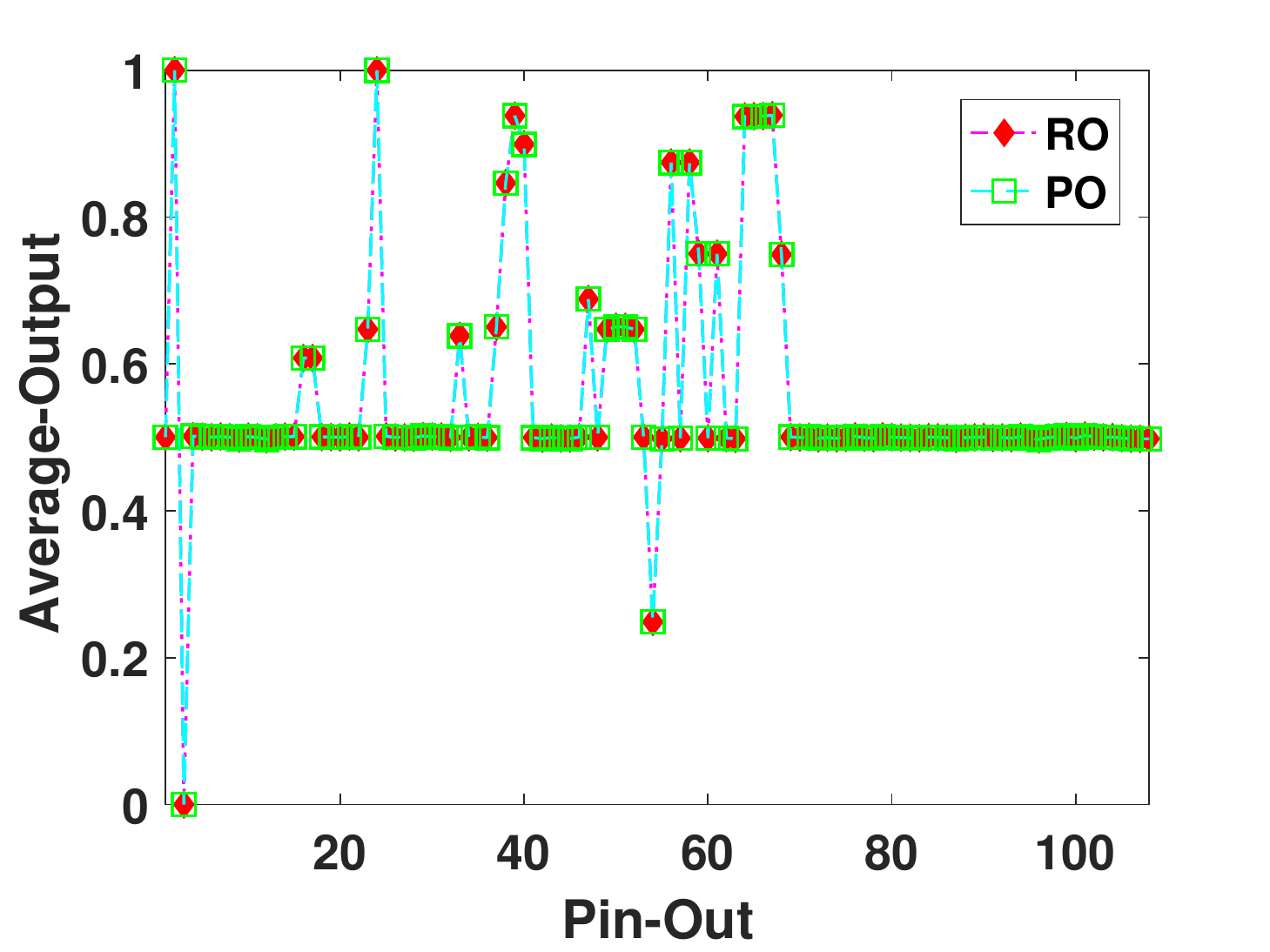} \\
(a) & (b) & (c) & (d)
\end{tabular}
\caption{Average prediction of output patterns from four ISCAS benchmarks (a) c423, (b) c1355, (c) c1908, and (d) c7552.}
\label{output-guessing}
\end{figure*}

\begin{figure*}
\begin{tabular}{cccc}
\includegraphics[width=0.23\linewidth]{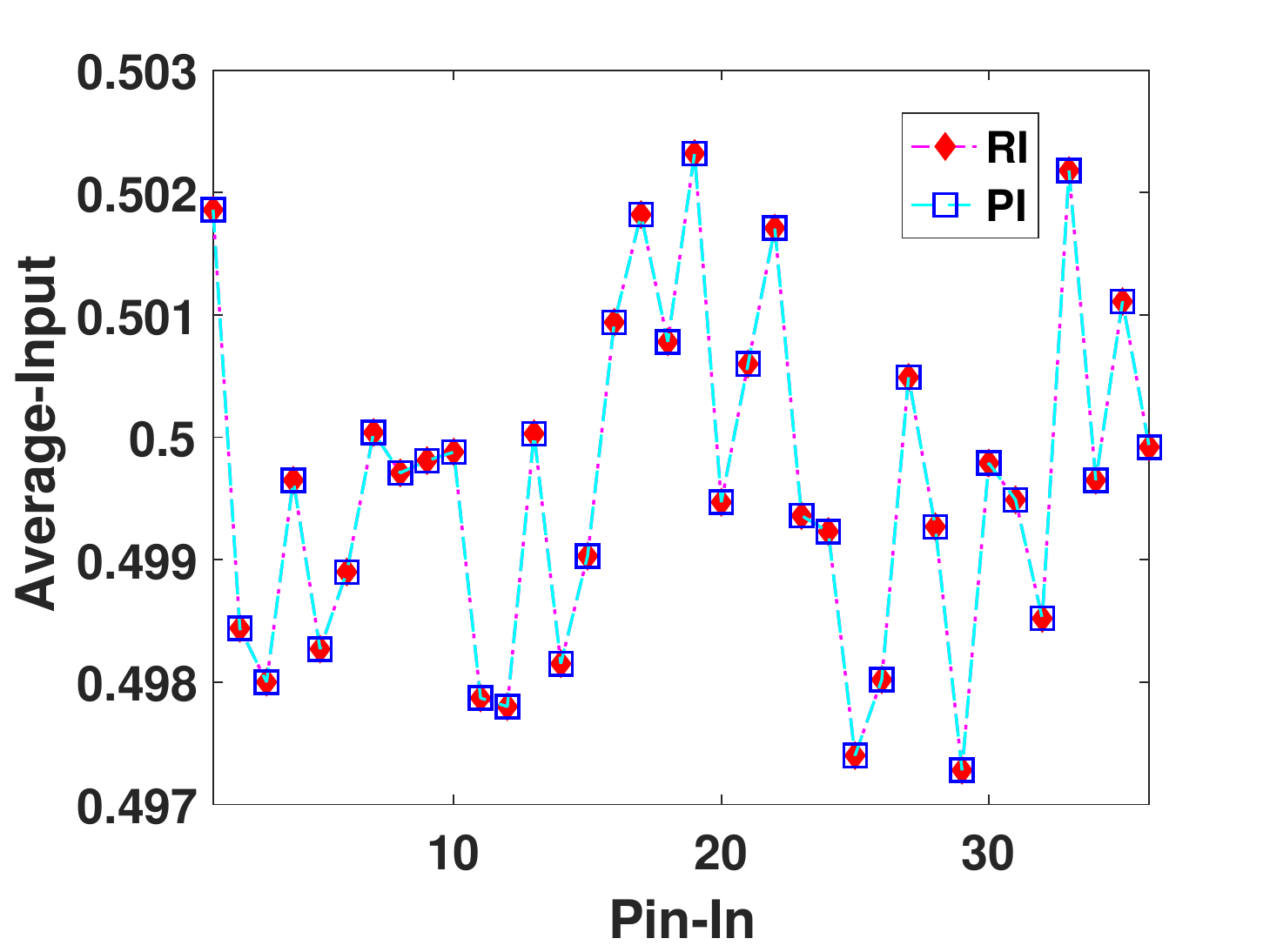} &
\includegraphics[width=0.23\linewidth]{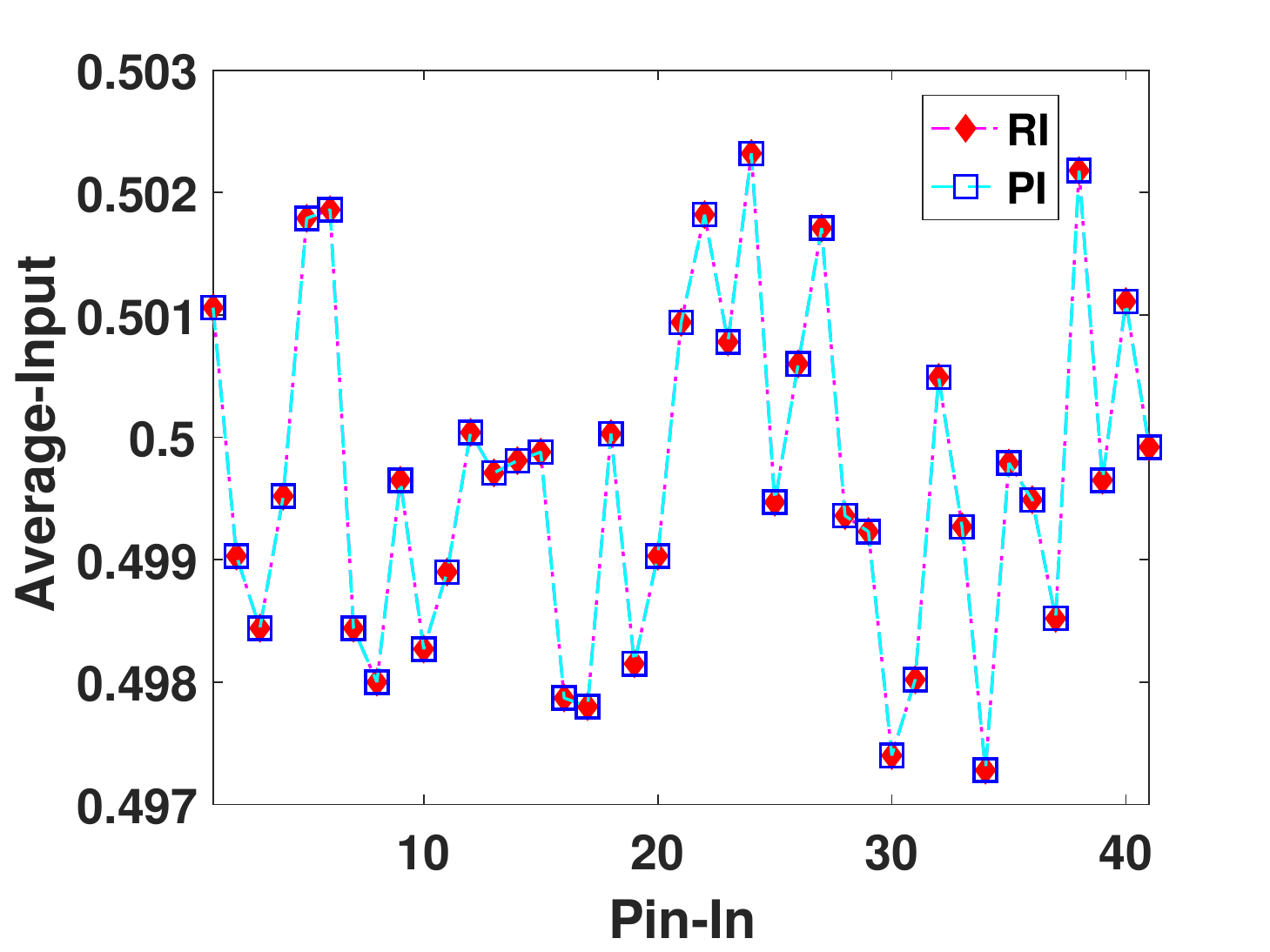} &
\includegraphics[width=0.23\linewidth]{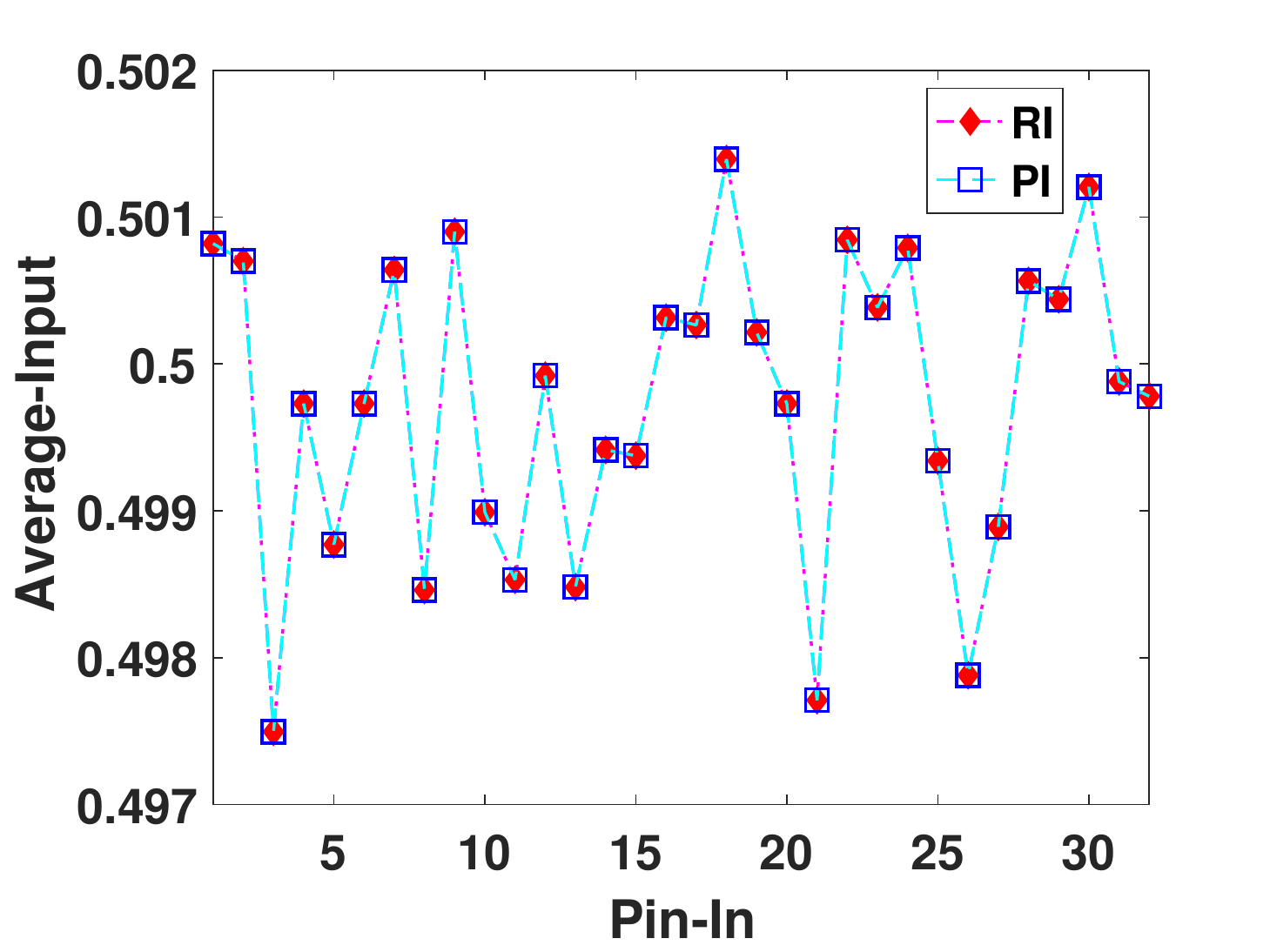} &
\includegraphics[width=0.23\linewidth]{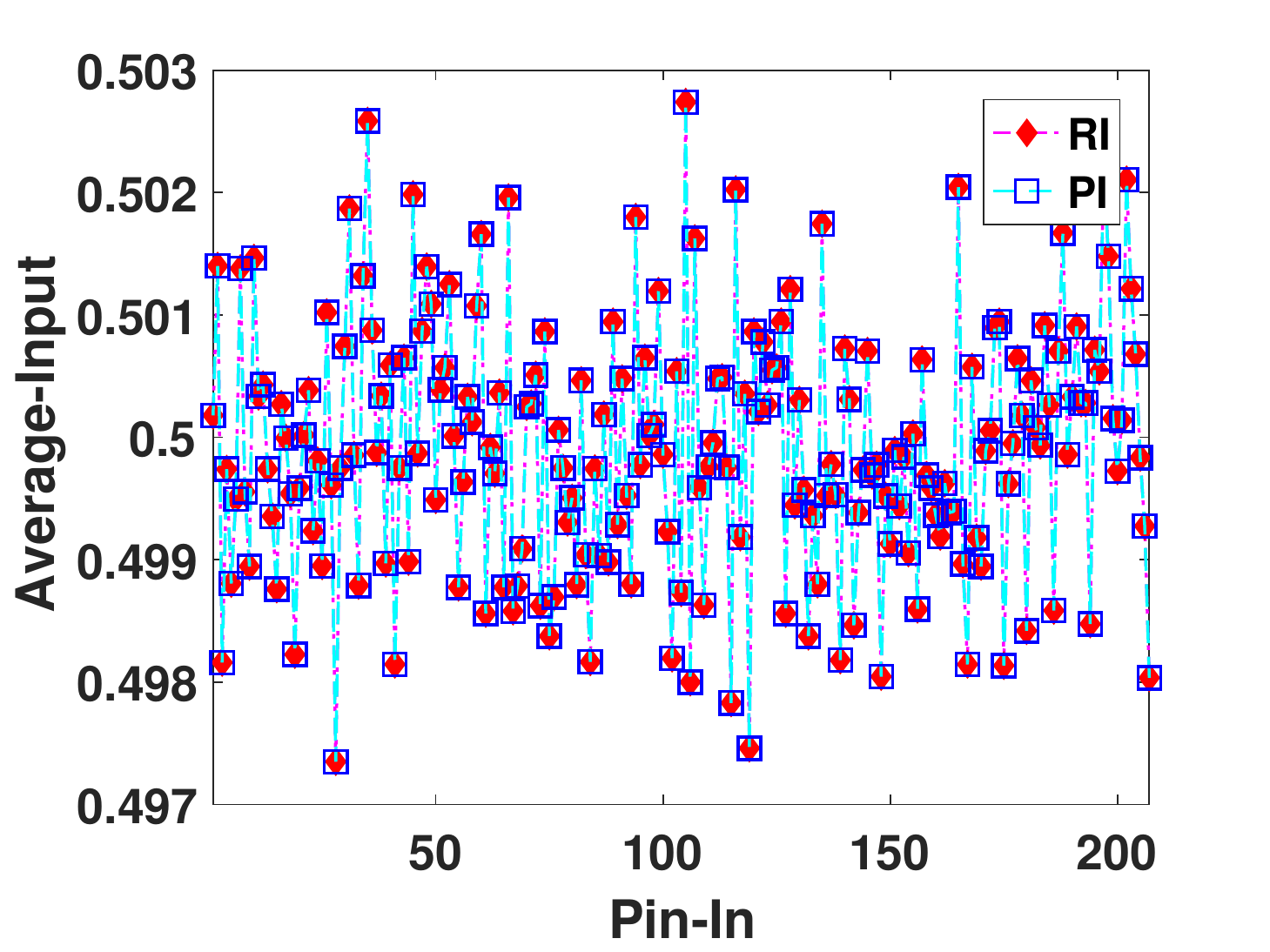} \\
(a) & (b) & (c) & (d)
\end{tabular}
\caption{Average prediction of input patterns for four ISCAS benchmarks (a) c423, (b) c1355, (c) c1908, and (d) c7552.}
\label{input-guessing}
\end{figure*}

\subsubsection{\textbf{Attack2-Output Guessing}}
This attack scenario has two phases: i) Training phase: During the training phase, the D-RNN is supervised to learn from the data by presenting the training data at the input layer and dynamically adjusting the parameters of the D-RNN to achieve the desired output value from the input set. Basically, an attacker trains I/O pairs to train D-RNN parameters. Once the parameters are trained, by giving new input patterns, D-RNN will be able to predict the corresponding output stimulus. In this scenario, we train only less than half a percent (< 0.5\%) of the total available I/O patterns. ii) Testing phase: In order to evaluate our attack model, we test D-RNN with one million new untrained input patterns to predict the outputs. Next, we compare the predicted outputs with the desired ones to ensure the success of our attack model. This scenario is applied to all 4 different ISCAS-85 benchmarks with various number of inputs and outputs. The results of our work is shown in Fig.~\ref{output-guessing}. In this figure, x-axis and y-axis show the number of outputs (pin-out) and average value of the corresponding output pin, respectively. Note that average value here is the average possible outputs from each benchmark. Moreover, RO (red line) indicates ``real output'' and PO (green line) denotes the ``predicted'' output from Deep-RNN. Fig.~\ref{output-guessing}(a) is the result for benchmark1 (ISCAS-85 c423) that clearly demonstrates that these two lines (RO and PO) are overlapped; meaning D-RNN could predict the output patterns for this benchmark successfully. Similar scenario is seen for Figs.~\ref{output-guessing}(b-d) which obviously have more output patterns to be predicted from benchmark2 (ISCAS-85 c1355), benchmark3 (ISCAS-85 c1908), and benchmark4 (ISCAS-85 c7552). Note that in this attack scenario, D-RNN does not need a secret key to predict output patterns.

\subsubsection{\textbf{Attack3-Input Guessing}}
This attack model is very similar to the concepts that we have defined in Attack2 except that the inputs will be predicted by considering recursive D-RNN. In this scenario, an attacker applies desired outputs to predict the input stimuli. The results of our work are shown in Fig.~\ref{input-guessing}. In this figure, x-axis and y-axis show the number of inputs (pin-in) and average value of the corresponding input pin, respectively. Similar to the previous attack, Fig.~\ref{input-guessing}(a) is the result for benchmark1 (ISCAS-85 c423) demonstrating that these two lines (RI and PI) are overlapped; meaning D-RNN could predict the input patterns for this benchmark successfully. Similar scenario is seen for Figs.~\ref{input-guessing}(b-d) (have more input patterns to be predicted from benchmark2 (ISCAS-85 c1355), benchmark3 (ISCAS-85 c1908), and benchmark4 (ISCAS-85 c7552)). Similar to Attack2 scenario, in this attack model, D-RNN does not need a secret key to predict the output patterns.

\section{\textbf{CONCLUSION AND FUTURE WORK}}
This paper, for the first time, explores the potential of employing the state-of-the-art deep learning technique (Deep RNN) that allows an attacker to derive the correct values of the key inputs (secret key) from logic encryption hardware obfuscation techniques. We have also presented ``output-guessing" and ``input-guessing" attacks in which an attacker can guess outputs and inputs without having the knowledge of secret key, respectively. Our result indicates that the proposed method can deliver significant improvement for breaking logic encryption. Compared with state-of-the-art attacks, the proposed method (BOCANet) is 20 times faster with relatively-high success rate using only a small number of input/output observations (0.5\%) taken from an activated IC. We believe that with augmenting hardware attacks on obfuscated architectures by incorporating deep learning in our scheme not only a practical paradigm shift to attack such obscured circuits is proposed but also it can be utilized as a step forward towards finding the vulnerabilities of other hardware security approaches. As future work, we will investigate (i) if the proposed approach would make delay+logic locking (DLL) based countermeasures more effective and (ii) if the presented scheme could be augmented by such countermeasures to introduce even more powerful mechanisms. We also intend to develop side-channel attack-based mechanisms, e.g., through first and higher order differential power analysis attacks, to make the proposed schemes, potentially, even more powerful.


%
%
%



%
%
%
%

\bibliographystyle{IEEEtran}
\bibliography{ieeeREF}

\end{document}